\documentclass[aps,prl,reprint,superscriptaddress]{revtex4-2}

\usepackage{textcomp}
\usepackage{array}

\usepackage{graphicx}
\usepackage{float}
\usepackage{amsmath}
\usepackage{amssymb}
\usepackage{epstopdf}
\usepackage{bm}

\usepackage{siunitx}
\usepackage{natbib}

\usepackage{comment}
\usepackage{bold-extra}

\begin{document}

\title{InSbAs two-dimensional electron gases as a platform for topological superconductivity}

\author{Christian~M.~Moehle$^{*}$}
\affiliation{QuTech and Kavli Institute of Nanoscience, Delft University of Technology, 2600 GA Delft, The Netherlands}

\author{Chung~Ting~Ke$^{*}$}
\affiliation{QuTech and Kavli Institute of Nanoscience, Delft University of Technology, 2600 GA Delft, The Netherlands}

\author{Qingzhen~Wang}
\affiliation{QuTech and Kavli Institute of Nanoscience, Delft University of Technology, 2600 GA Delft, The Netherlands}

\author{Candice~Thomas}
\affiliation{Department of Physics and Astronomy, Purdue University, West Lafayette, Indiana 47907, USA}
\affiliation{Birck Nanotechnology Center, Purdue University, West Lafayette, Indiana 47907, USA}

\author{Di~Xiao}
\affiliation{Department of Physics and Astronomy, Purdue University, West Lafayette, Indiana 47907, USA}
\affiliation{Birck Nanotechnology Center, Purdue University, West Lafayette, Indiana 47907, USA}

\author{Saurabh~Karwal}
\affiliation{QuTech and Netherlands Organization for Applied Scientific Research (TNO), 2628 CK Delft, The Netherlands}

\author{Mario~Lodari}
\affiliation{QuTech and Kavli Institute of Nanoscience, Delft University of Technology, 2600 GA Delft, The Netherlands}

\author{Vincent~van~de~Kerkhof}
\affiliation{QuTech and Kavli Institute of Nanoscience, Delft University of Technology, 2600 GA Delft, The Netherlands}

\author{Ruben~Termaat}
\affiliation{QuTech and Kavli Institute of Nanoscience, Delft University of Technology, 2600 GA Delft, The Netherlands}

\author{Geoffrey~C.~Gardner}
\affiliation{Microsoft Quantum Purdue, Purdue University, West Lafayette, Indiana 47907, USA}

\author{Giordano~Scappucci}
\affiliation{QuTech and Kavli Institute of Nanoscience, Delft University of Technology, 2600 GA Delft, The Netherlands}

\author{Michael~J.~Manfra}
\affiliation{Department of Physics and Astronomy, Purdue University, West Lafayette, Indiana 47907, USA}
\affiliation{Birck Nanotechnology Center, Purdue University, West Lafayette, Indiana 47907, USA}
\affiliation{School of Electrical and Computer Engineering, Purdue University, West Lafayette, Indiana 47907, USA}
\affiliation{Microsoft Quantum Purdue, Purdue University, West Lafayette, Indiana 47907, USA}
\affiliation{School of Materials Engineering, Purdue University, West Lafayette, Indiana 47907, USA}

\author{Srijit~Goswami$^{+}$}
\affiliation{QuTech and Kavli Institute of Nanoscience, Delft University of Technology, 2600 GA Delft, The Netherlands}

\begin{abstract}
$^{+}$ E-mail: S.Goswami@tudelft.nl\\	
	
Topological superconductivity can be engineered in semiconductors with strong spin-orbit interaction coupled to a superconductor. Experimental advances in this field have often been triggered by the development of new hybrid material systems. Among these, two-dimensional electron gases (2DEGs) are of particular interest due to their inherent design flexibility and scalability. Here we discuss results on a 2D platform based on a ternary 2DEG (InSbAs) coupled to in-situ grown Aluminum. The spin-orbit coupling in these 2DEGs can be tuned with the As concentration, reaching values up to $400$\,$\mathrm{meV\AA}$, thus exceeding typical values measured in its binary constituents. In addition to a large Landé g-factor $\sim$ 55 (comparable to InSb), we show that the clean superconductor-semiconductor interface leads to a hard induced superconducting gap. Using this new platform we demonstrate the basic operation of phase-controllable Josephson junctions, superconducting islands and quasi-1D systems, prototypical device geometries used to study Majorana zero modes.\\

\end{abstract}

\maketitle

Topological phases of matter are currently a subject of intense research. Following early theoretical proposals~\cite{Lutchyn_2010, Oreg_2010}, materials with large spin-orbit interaction (such as InAs and InSb) coupled to superconductors have emerged as a promising platform to engineer topological superconductivity in the form of Majorana zero modes (MZMs). In this context, one-dimensional nanowires have been studied extensively over the years~\cite{Aguado_2017, Lutchyn_2018, Prada_2020}. More recently, several efforts have been focused on engineering MZMs in two-dimensional electron gases (2DEGs). Not only do 2DEGs provide a scalable platform for future development of topological qubits, but their inherent flexibility allows for the realization of more complex devices. The versatility of the 2DEG platform can be seen in the variety of experiments performed on quasi-1D structures~\cite{Nichele_2017}, superconducting islands~\cite{Farrell_2018}, multi-terminal Josephson junctions (JJs)~\cite{Pankratova_2020}, and phase-biased JJs~\cite{Fornieri_2019, Ren_2019}, all of which are promising architectures to create topological systems. Many of these studies have been performed on InAs 2DEGs where it is possible to create a pristine interface between the superconductor Aluminum (Al) and the 2DEG allowing for a strong superconducting proximity effect~\cite{Shabani_2016, Kjaergaard_2016}.

The InSb 2DEG is another appealing platform, primarily due to its significantly larger g-factor and spin-orbit coupling. Whereas the former allows the hybrid system to enter the topological regime at a lower magnetic field, the latter is crucial in determining the topological gap that protects the MZMs. These 2DEGs have recently been proximitized by ex-situ superconductors~\cite{Ke_2019}, however there exist no reports of InSb-based hybrid systems with in-situ grown superconductors. This could be related to the band offset at the InSb-Al interface, which (unlike InAs) prevents an efficient accumulation of charge carriers and hence induced superconductivity~\cite{Schuwalow_2019}. It would thus be ideal to have a material system with the desirable properties of both InAs and InSb.

In this work we explore such a new hybrid material: ternary (InSbAs) 2DEGs coupled to in-situ grown Al.  Using magneto-transport experiments we demonstrate a large g-factor ($\sim$ 55) and exceptionally strong spin-orbit coupling exceeding the values of either InAs or InSb. In addition, the pristine semiconductor-superconductor interface leads to a hard induced superconducting gap that is revealed by spectroscopy measurements. Furthermore, using these ternary 2DEGs we demonstrate the stable operation of prototypical devices studied in the context of MZMs: phase-controllable JJs, superconducting islands, and quasi-1D structures. Our results show that \mbox{InSbAs/Al} 2DEGs offer the combined advantages of their binary constituents and are therefore a promising platform to realize topological superconductivity.

\begin{figure}[!t]
	\centering
	\includegraphics[scale=0.8]{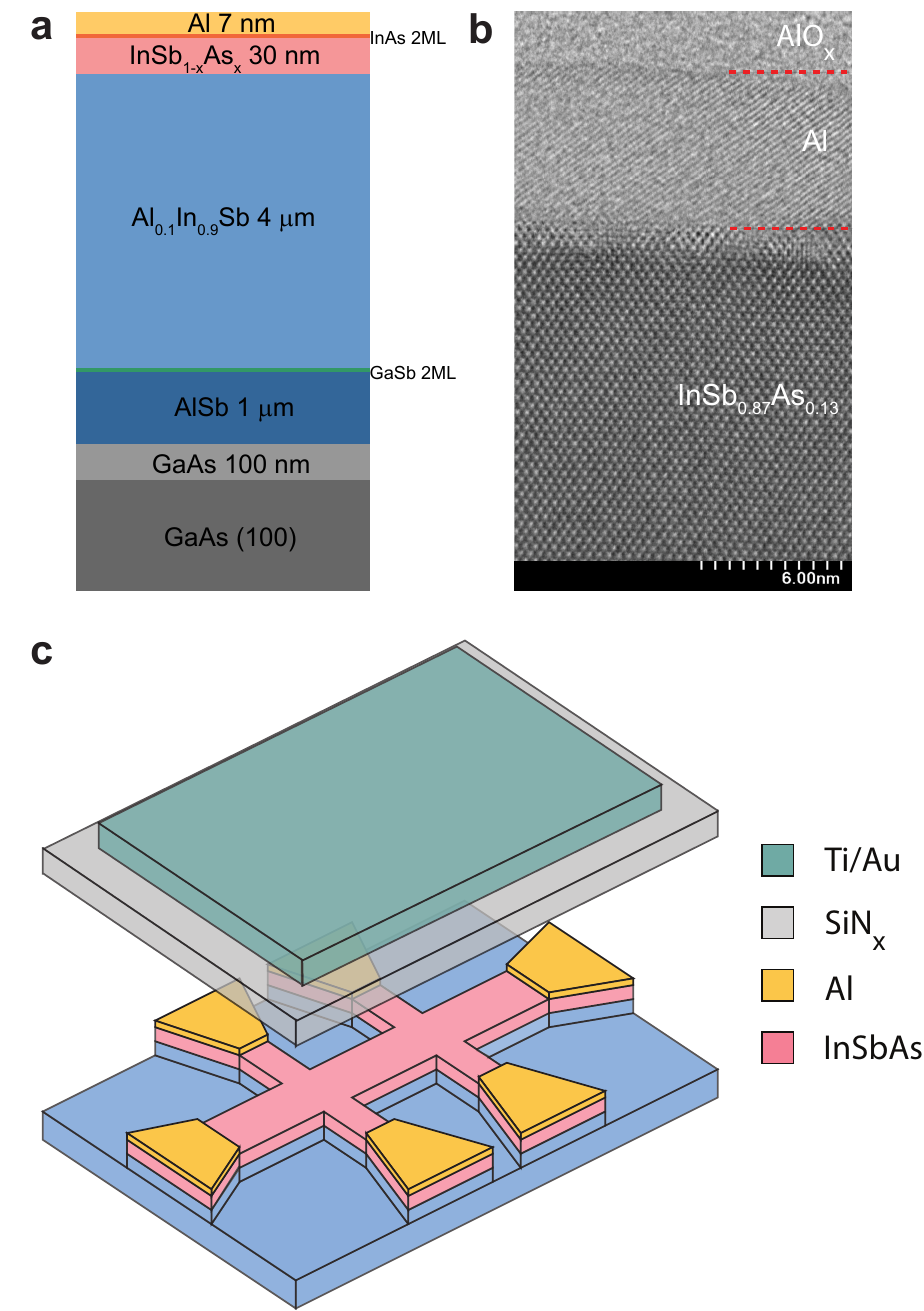}\\
	\caption{\textbf{Hybrid Al/InSbAs heterostructures.} \textbf{a,} Layer stack of the Al-$\mathrm{InSb}_{1-x}\mathrm{As}_{x}$ hybrid heterostructures. \textbf{b,} Bright-field scanning transmission electron micrograph of the Al-$\mathrm{InSb_{0.870}As_{0.130}}$ interface along the [110] zone axis of the semiconductor. Red lines indicate the boundaries of the Aluminum. \textbf{c,} Schematic of a Hall bar that is used to extract the 2DEG properties.}
	\label{fig1}
\end{figure}

$\mathrm{InSb}_{1-x}\mathrm{As}_{x}$ 2DEGs with varying As concentration, $x$, are grown by molecular beam epitaxy (MBE) on undoped, semi-insulating GaAs (100) substrates (see Fig. \ref{fig1}a for a schematic of the layer stack). The growth starts with a \SI{100}{nm} GaAs buffer layer, directly followed by a \SI{1}{\mu m} thick AlSb nucleation layer~\cite{Goldammer_1999} and a \SI{4}{\mu m} thick $\mathrm{Al_{0.1}In_{0.9}Sb}$ layer. The latter forms a closely matched pseudo-substrate for the $\mathrm{InSb}_{1-x}\mathrm{As}_{x}$ growth and the bottom barrier of the quantum well~\cite{Lehner_2018}. The As concentration in the $\mathrm{InSb}_{1-x}\mathrm{As}_{x}$ is controlled by the growth temperature and the As flux. In this study, heterostructures with $x=0$, 0.053, 0.080, 0.130, 0.140 and 0.240 are grown. The semiconductor growth is terminated by the deposition of 2 monolayers (ML) InAs, serving as a screening layer to prevent intermixing between the semiconductor structure and the superconducting Al layer~\cite{Thomas_2019}. After the semiconductor growth, the heterostructures are transferred under ultra-high vacuum to a second MBE chamber to deposit \SI{7}{nm} of Al, using methods described in~\cite{Thomas_2019}. Figure \ref{fig1}b displays a bright field scanning transmission electron micrograph focusing on the Al/$\mathrm{InSb}_{1-x}\mathrm{As}_{x}$ interface for $x=0.130$. The interface appears sharp with a slight change of atomic contrast that is attributed to the relaxed InAs screening layer~\cite{Thomas_2019}. Further details about the growth process can be found in the Supplementary Information (SI).

We characterize the semiconducting properties of the InSbAs 2DEGs by removing the Al in the active device area to fabricate Hall bars. After the Al removal, the 2DEG is etched in unwanted areas, followed by the deposition of a $\mathrm{SiN}_{x}$ dielectric layer. Lastly, a Ti/Au top-gate is evaporated and used to control the electron density in the 2DEG (see Fig. \ref{fig1}c for a schematic). We find peak mobilities of $\SI{20000}-\SI{28000}{cm^2/Vs}$ (see SI for mobility-density curves, and further details about the device fabrication).

\begin{figure*}[!t]
	\includegraphics[width=1.0\textwidth]{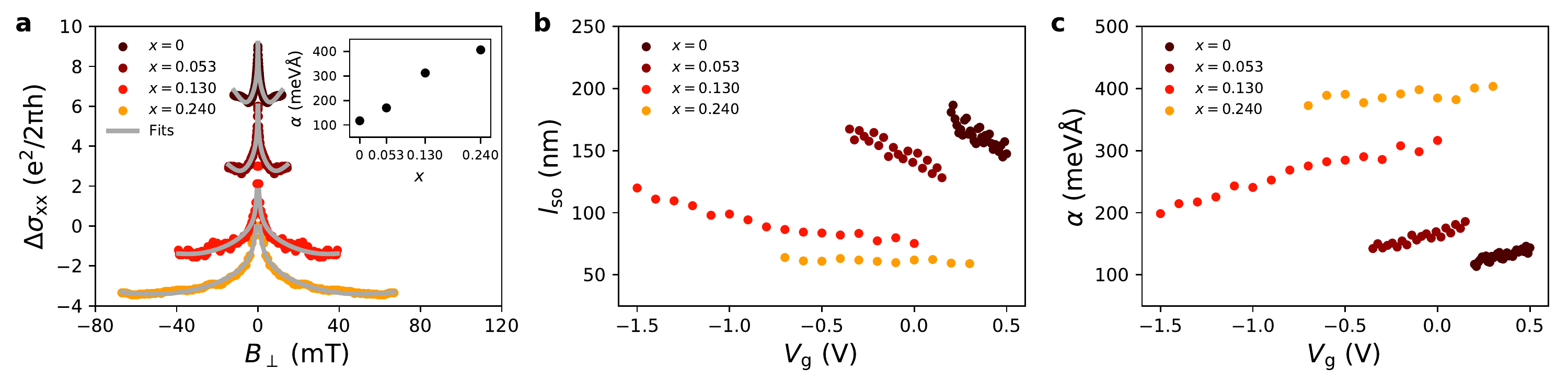}
	\caption{\textbf{Large and tuneable spin-orbit coupling.} \textbf{a,} Magneto-conductivity correction at $V_{\mathrm{g}}=\SI{0}{V}$ for the different $\mathrm{InSb}_{1-x}\mathrm{As}_{x}$ 2DEGs. The $x=0$ curve is measured at $V_{\mathrm{g}}=\SI{0.2}{V}$ due to a high resistance at $V_{\mathrm{g}}=\SI{0}{V}$. The gray lines are ILP fits to the weak anti-localization data. In the inset the extracted linear Rashba coefficient is plotted for the four As concentrations, showing a monotonic increase with increasing As concentration. For the higher As concentrations, $\alpha$ is 3-4 times larger as compared to the value for pure InSb ($x=0$). \textbf{b,} Spin-orbit length plotted against $V_{\mathrm{g}}$. \textbf{c,} $\alpha$ as a function of $V_{\mathrm{g}}$. $l_{\mathrm{so}}$ ($\alpha$) decreases (increases) with increasing As concentration when comparing at a fixed gate voltage.}
	\label{fig3}
\end{figure*}

To study the spin-orbit coupling in these 2DEGs, we measure the longitudinal conductivity, $\sigma_{\mathrm{xx}}$, in perpendicular magnetic fields, $B_{\perp}$, at $\SI{300}{mK}$ using standard lock-in techniques. The simultaneously measured transversal Hall resistance allows us to deduce the density, $n$, in the 2DEG at every gate voltage, $V_{\mathrm{g}}$. Figure \ref{fig3}a shows the magneto-conductivity correction, $\Delta \sigma_{\mathrm{xx}}(B_{\perp})=\sigma_{\mathrm{xx}}(B_{\perp})-\sigma_{\mathrm{xx}}(0)$, at $V_{\mathrm{g}}=\SI{0}{V}$ for $\mathrm{\mathrm{x}=0}$, 0.053, 0.130 and 0.240, where the individual curves are offsetted for clarity. We observe clear weak anti-localization (WAL) peaks that are caused by the suppression of coherent backscattering due to spin-orbit coupling. Since we expect the linear Rashba term to be the dominating spin-orbit contribution in these asymmetric quantum wells, we fit the WAL peaks with the Rashba-dominated Iordanskii, Lyanda-Geller, and Pikus (ILP) model~\cite{Iordanskii_1994, Knap_1996} (grey curves in Fig. \ref{fig3}a). This allows us to extract the spin-orbit length, $l_{\mathrm{so}}=\sqrt{D\tau_{\mathrm{so}}}$, where $D=v_{\mathrm{F}}l_{\mathrm{e}}/2$ is the diffusion constant. Here, $\tau_{\mathrm{so}}$ is the spin-orbit scattering time, $v_{\mathrm{F}}$ the Fermi velocity and $l_{\mathrm{e}}$ the mean free path. The linear Rashba parameter is given by $\alpha=\Delta_{\mathrm{so}}/2k_{\mathrm{F}}$, where $\Delta_{\mathrm{so}}=\sqrt{2\hbar^2/\tau_{\mathrm{so}}\tau_{\mathrm{e}}}$ is the spin-split energy, $k_{\mathrm{F}}$ the Fermi wave vector, and $\tau_{\mathrm{e}}$ the elastic scattering time.

As shown in Fig. \ref{fig3}a, the ILP model fits the experimental data well. The resulting linear Rashba parameter for the four different As concentrations is plotted in the inset. It is striking that $\alpha$ increases monotonically with increasing As concentration. Compared to the value for pure InSb ($\alpha\approx\SI{100}{meV\AA}$), it is noteworthy that for the higher As concentrations, the linear Rashba parameter is 3-4 times larger ($300-400$\,$\mathrm{meV\AA}$). We proceed by measuring WAL as a function of gate voltage ($V_{\mathrm{g}}$) for the different As concentrations. In Fig. \ref{fig3}b and \ref{fig3}c we show $l_{\mathrm{so}}$ and $\alpha$ plotted against $V_{\mathrm{g}}$ (see SI for the same plots as a function of electron density). The trend of decreasing (increasing) $l_{\mathrm{so}}$ ($\alpha$) with increasing As concentration persists also when comparing at other gate voltages. We note, however, that specifically for $x=0.240$, the spin-orbit coupling becomes so large that $l_{\mathrm{so}}$ is smaller than $l_{\mathrm{e}}$. This might lead to inaccuracies in the extracted fit parameters as the ILP model is valid when $l_{\mathrm{e}}$ is the smallest length scale.

The systematic increase in spin-orbit coupling with As concentration can arise from a combination of several effects. Firstly, bandstructure calculations of $\mathrm{InSb}_{1-x}\mathrm{As}_{x}$ show that the Rashba parameter is strongly influenced by the As concentration~\cite{Winkler_2016, Mayer_2020}, which has been observed in experiments on ternary nanowires~\cite{Sestoft_2018}. Secondly, electric fields across the 2DEG can also influence the spin-orbit interaction. We note that even at $V_{\mathrm{g}}=\SI{0}{V}$, $\alpha$ increases monotonically with $x$, suggesting that the external electric field from the applied gate voltage is not the primary source of the enhancement. However the internal field (generated at the 2DEG-gate dielectric interface) could be a strong function of the As concentration. We indeed observe that the nominal density in the 2DEG (density at $V_{\mathrm{g}}=\SI{0}{V}$) increases systematically with As concentration (see SI), indicating a stronger downward band bending at the dielectric-semiconductor interface resulting in an increased electric field. While from these studies it is difficult to disentangle the effects of the bulk semiconductor from the interfaces, similar experiments on deep InSbAs quantum wells would shed more light on the origins of the enhanced spin-orbit interaction.  

Having established strong spin-orbit coupling in $\mathrm{InSb}_{1-x}\mathrm{As}_{x}$ 2DEGs, we proceed by measuring the perpendicular g-factor, $g^*$. By comparing the temperature dependence of Shubnikov-de Haas oscillations for an odd-even filling factor couple, an expression for $g^*$ can be obtained~\cite{Drichko_2018}. For $x=0$ we find $g^*=47.8\pm 2.8$. The same analysis is done for $x=0.130$, where we obtain $g^*=54.6\pm 3.1$ (see SI for details about the data analysis, as well as effective mass measurements for the two As concentrations). This shows that besides strong spin-orbit coupling, InSbAs 2DEGs also possess a large g-factor that is comparable to the one of pure InSb.  

\begin{figure*}[!t]
	\includegraphics[width=1.0\textwidth]{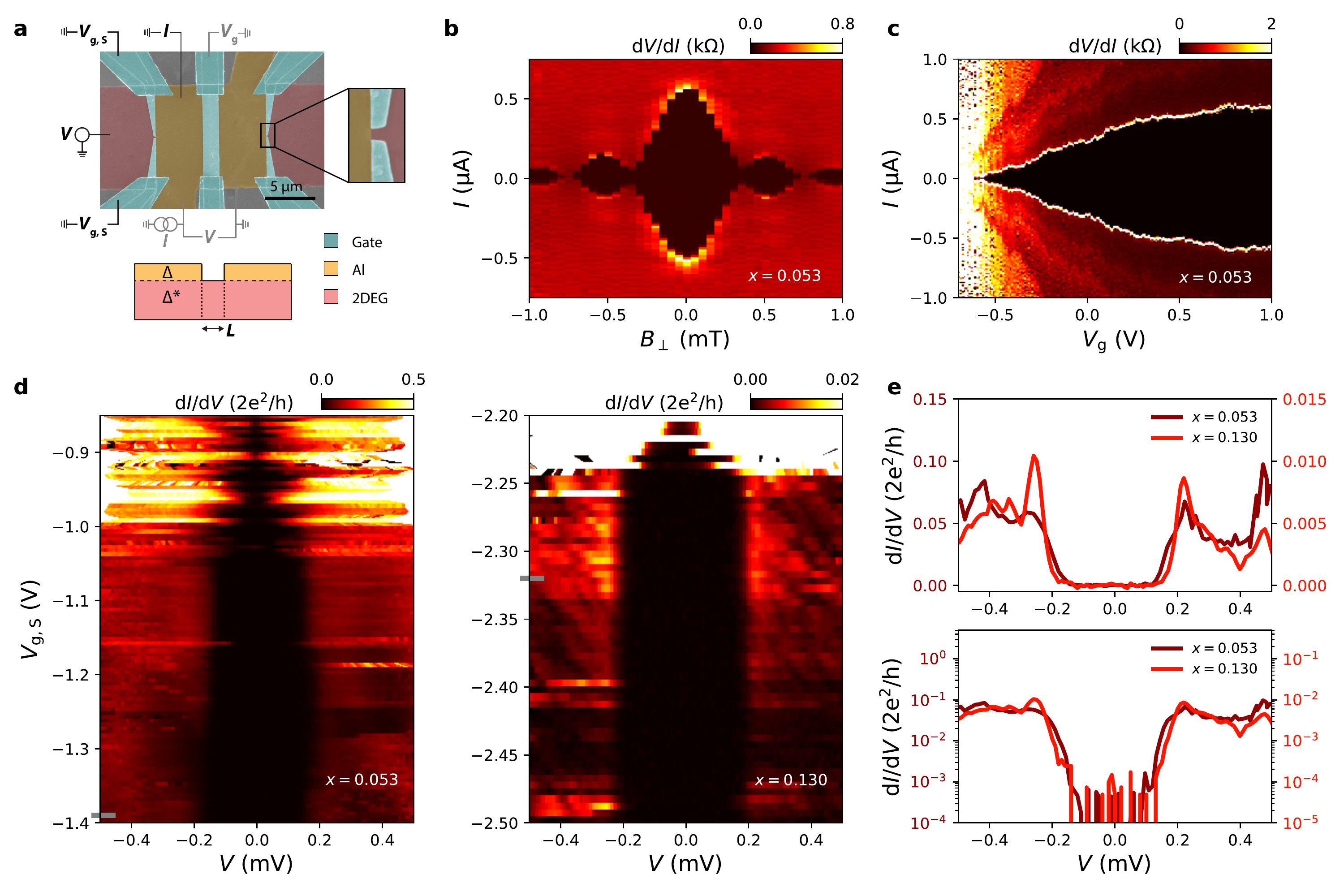}
	\caption{\textbf{Hard induced superconducting gap.} \textbf{a,} False-colored SEM of a combined JJ and tunneling spectroscopy device. A cross-sectional schematic the JJ part is shown in the bottom. All devices have a fixed JJ length of $L\approx\SI{150}{nm}$. \textbf{b,} Differential resistance as a function of applied current bias and perpendicular magnetic field for the $x=0.053$ JJ. The Fraunhofer interference pattern signifies a uniform current distribution in the JJ. \textbf{c,} Differential resistance as a function of applied current bias and gate voltage for the same JJ, showing that the switching current, $I_{\mathrm{s}}$, can be fully suppressed. \textbf{d,} Differential conductance as a function of splitgate voltage and applied voltage bias for $x=0.053$ and $x=0.130$. The color scale has been saturated to increase the visibility of the tunneling regime. For both As concentrations, the induced superconducting gap is visible (region of suppressed conductance), stable over a large range in $V_{\mathrm{g, S}}$. \textbf{e,} Linecuts at the indicated positions (gray markers) on linear and logarithmic scale. The size of the induced gap is similar for both As concentrations. The in-gap conductance is suppressed by 2-3 orders of magnitude as compared to the out-of gap conductance, indicating a hard induced gap.}
	\label{fig4}
\end{figure*}

Given that the semiconducting properties of InSbAs 2DEGs are favorable to realize topological phases, we now demonstrate that these 2DEGs also have excellent coupling to Al. To do so, we use devices as shown in the false-colored scanning electron micrograph (SEM) in Fig. \ref{fig4}a. This device can either be operated as a gate-tunable JJ (gray circuit) or a spectroscopy device (black circuit) to measure the local density of states. All devices are measured at a temperature of \SI{30}{mK}.

It is important to note that we do not see any induced superconductivity for pure InSb, presumably due to an unfavorable band alignment at the Al-InSb interface. In stark contrast, all the InSbAs 2DEGs (irrespective of As concentration) have excellent coupling to the superconductor, where all JJs display supercurrents and pronounced multiple Andreev reflections (see SI). In Fig. \ref{fig4}b we show a representative Fraunhofer interference pattern for the JJ with $x=0.053$, where the black regions correspond to zero resistance. The size of the switching current, $I_{\mathrm{s}}$, can be controlled by the gate voltage, as demonstrated in Fig. \ref{fig4}c for the same As concentration. Upon lowering $V_{\mathrm{g}}$, $I_{\mathrm{s}}$ shrinks and correspondingly, the normal-state resistance of the JJ, $R_{\mathrm{n}}$, increases. 

Clean and transparent interfaces are crucial for the realization of MZMs as they allow the proximitized semiconductor to obtain a hard induced superconducting gap with a vanishing in-gap density of states. It is therefore important to measure the density of states in the proximitized 2DEG directly. To this end, we operate the devices shown in Fig. \ref{fig4}a as spectroscopy devices. We apply a voltage bias to the left normal contact, and measure the current flowing through the Al lead, while energizing the splitgates with a negative voltage, $V_{\mathrm{g, S}}$, to create a barrier. In the tunneling regime the measured conductance is proportional to the density of states in the proximitized 2DEG. In Fig. \ref{fig4}d we present tunneling spectroscopy maps for $x=0.053$ and $x=0.130$. In both measurements we note the emergence of a region with suppressed conductance, reflecting the induced superconducting gap. The gaps persist over an extended range in $V_{\mathrm{g, S}}$ with a slight dependence of the gap size on the out-of-gap conductance. In Fig. \ref{fig4}e we show two representative linecuts in the tunneling regime. We find that the size of the induced gap is similar for both As concentrations ($\Delta^{*}\approx\SI{220}{\mu eV}$). Turning to the linecuts on the logarithmic scale and comparing the out-of-gap conductance with the in-gap conductance, we see that the in-gap conductance is suppressed by 2-3 orders of magnitude for both As concentrations. This confirms the excellent Al-2DEG interface.\\

\begin{figure*}[!t]
	\includegraphics[width=1.0\textwidth]{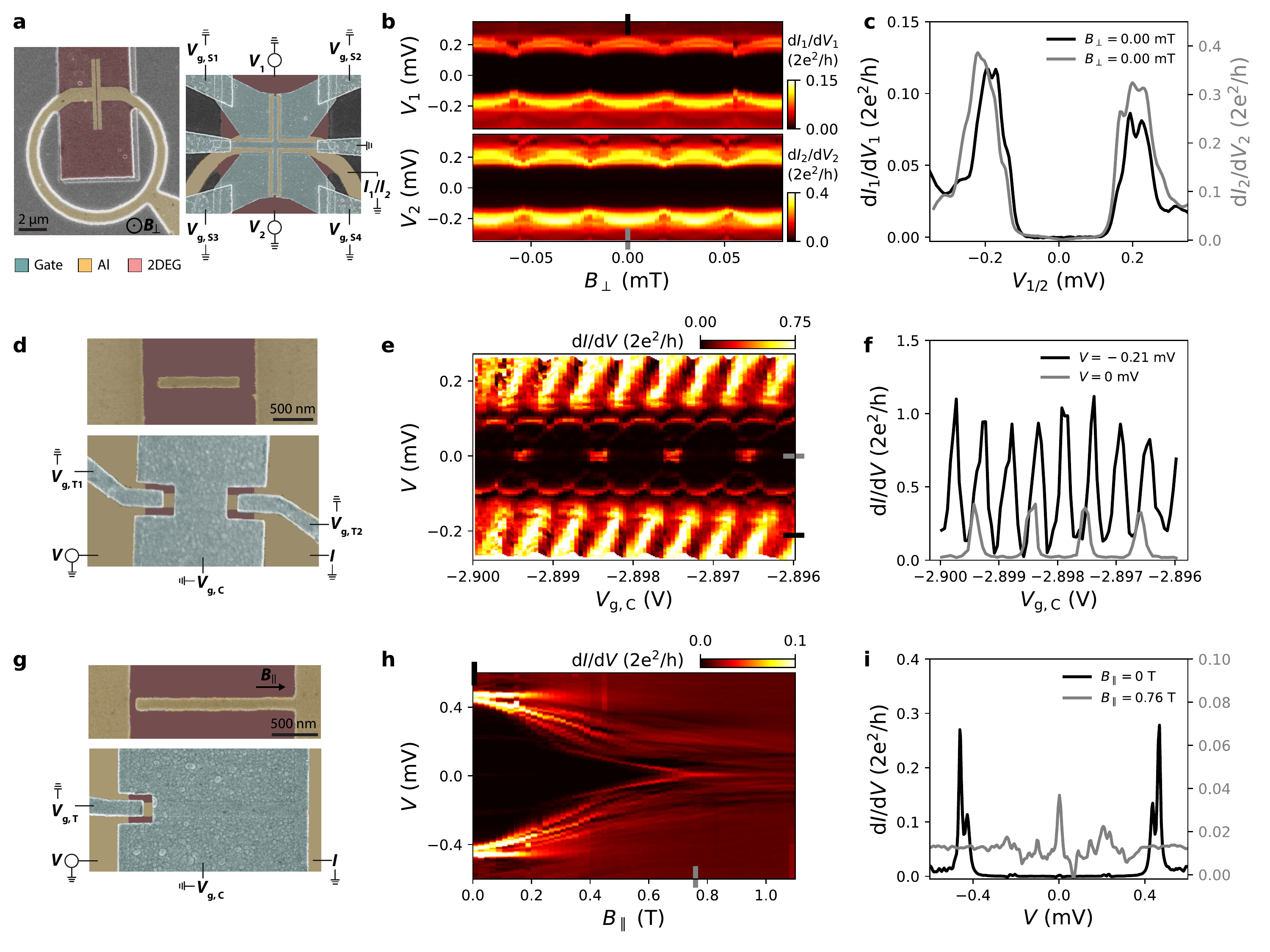}
	\caption{\textbf{Prototypical Majorana devices.} \textbf{a,} False-colored SEMs of a phase-biased JJ before (left) and after (right) gate deposition. The split gates at the top and bottom edge of the JJ are used to perform tunneling spectroscopy at both ends of the JJ. This device is fabricated on a 2DEG with $x=0.080$. \textbf{b,} Differential conductance as a function of voltage bias applied to the top contact and perpendicular magnetic field while the bottom tunneling contact is floating (top panel). The bottom panel shows the tunneling spectroscopy map as a function of bottom voltage bias and perpendicular magnetic field (top tunneling contact floating). For both measurements the splitgates are fixed at $V_{\mathrm{g, S1}}=\SI{-2.36}{V}$, $V_{\mathrm{g, S2}}=\SI{-1.36}{V}$ and $V_{\mathrm{g, S3}}=V_{\mathrm{g, S4}}=\SI{-1.48}{V}$. \textbf{c,} Linecuts from \textbf{b} at the indicated positions. \textbf{d,} SEMs of a superconducting island before (top) and after (bottom) gate deposition, fabricated on a 2DEG with $x=0.140$. The two tunnel gates are used to tune the transmission between the leads and the island. The central gate depletes the surrounding 2DEG and changes the charge occupancy of the island. \textbf{e,} Differential conductance at fixed tunnel gate voltages ($V_{\mathrm{g, T1}}=\SI{-2.157}{V}$, $V_{\mathrm{g, T2}}=\SI{-2.052}{V}$) as a function of applied voltage bias and central gate voltage. 2e-periodic Coulomb oscillations are visible at $V=\SI{0}{mV}$ and 1e-periodic Coulomb oscillations at high biases. \textbf{f,} Linecuts from \textbf{e} at the indicated positions. \textbf{g,} SEMs of a quasi-1D grounded superconducting strip before (top) and after (bottom) gate deposition. The left gate is used to create a tunnel barrier between the bulk Al contact and the superconducting strip, and the central gate depletes the surrounding 2DEG. This device is fabricated on a 2DEG with $x=0.140$. \textbf{h,} Differential conductance at fixed gate voltages ($V_{\mathrm{g, T}}=\SI{-3.96}{V}$, $V_{\mathrm{g, C}}=\SI{-4.40}{V}$) as a function of applied voltage bias and parallel magnetic field. \textbf{i,} Linecuts from \textbf{h} at the indicated positions.}
	\label{fig6}
\end{figure*} 

Using the InSbAs/Al hybrid platform we realize three prototypical device architectures used to study MZMs: phase-controllable JJs, superconducting islands and quasi-1D superconducting strips. Figure \ref{fig6}a shows false-colored SEMs of a JJ embedded in a superconducting loop. A perpendicular magnetic field, $B_{\perp}$, penetrating through the loop can be used to tune the phase difference across the JJ. Split gates are positioned at the top and bottom edge of the JJ to perform tunneling spectroscopy at both ends of the phase-controllable JJ. In Fig. \ref{fig6}b we show spectroscopy maps at the top (top panel) and bottom (bottom panel) of the JJ, with representative linecuts shown in Fig. \ref{fig6}c. In both cases we observe a clear flux-dependent modulation of the gap, consistent with the phase modulation expected for Andreev bound states. Such three-terminal devices are important to check for correlations between the two ends of the JJ, when tuned into the topological regime.
   
A second approach to create MZMs is based on a floating narrow strip of superconductor, a superconducting island (see Fig. \ref{fig6}d). Here two tunnel gates are used to tune the transmission between the bulk Al contacts and the island. The central gate depletes the 2DEG in areas that are not covered with Al, and changes the charge occupancy on the island. In Fig. \ref{fig6}e we show a differential conductance map at fixed tunnel gate voltages, $V_{\mathrm{g,T}}$, varying the applied voltage bias, $V$, and the central gate voltage, $V_{\mathrm{g, C}}$. At $V=\SI{0}{mV}$ (see also gray linecut in Fig. \ref{fig6}f) we observe Coulomb peaks that have twice the spacing as compared to the Coulomb peaks at high biases (black linecut in Fig. \ref{fig6}f). The Coulomb peaks at $V=\SI{0}{mV}$ reflect 2e-periodic Cooper pair transport through the superconducting island, while at high biases, quasiparticles with 1e charge are allowed to tunnel through the island.

Another strategy to study MZMs employs a grounded narrow strip of superconductor (see Fig. \ref{fig6}g). The narrow gate is used to define a tunnel barrier between the bulk Al contact and the quasi-1D strip, and the central gate depletes the remaining exposed 2DEG. In Fig. \ref{fig6}h we present a tunneling spectroscopy map at fixed gate voltages as a function of applied voltage bias, $V$, and magnetic field, $B_{\parallel}$. $B_{\parallel}$ is oriented along the superconducting finger. Two representative linecuts are shown in Fig. \ref{fig6}i. Whereas at $B_{\parallel}=\SI{0}{T}$ we observe a hard induced superconducting gap (note that the gap size is doubled due to the superconducting contact), a zero energy state emerges around $B_{\parallel}=\SI{0.7}{T}$. While these results are promising, further experiments are required to comment on the origin of these states.     
  
In conclusion, we have shown that InSbAs 2DEGs offer the combined advantages of the more commonly studied binary materials. In addition to a large g-factor, they have excellent coupling to in-situ grown Aluminum. Furthermore, the spin-orbit coupling in these ternary 2DEGs is significantly stronger than in either InAs or InSb. Using this hybrid system, we realize distinct device architectures that can be used to study topological superconductivity.\\

\noindent
\textbf{\large Data availability}\\
Raw data and analysis scripts for all presented figures are available at the 4TU.ResearchData repository, DOI: https://doi.org/10.4121/14565228\\

\noindent
\textbf{\large Author information}\\
\noindent
\textbf{Corresponding author}\\
$^{+}$E-mail: S.Goswami@tudelft.nl\\

\noindent
\textbf{Author contributions}\\
$^{*}$C.~M.~M. and C.~T.~K contributed equally, fabricated and measured the devices together with Q.~W., S.~K. V.~vd~K. and R.~T. The MBE growth of the semiconductor heterostructures and the characterization of the materials was performed by C.~T., D.~X. and G.~C.~G. under the supervision of M.~J.~M. The effective mass and g-factor measurements were provided by M.~L. and G.~S. The manuscript was written by C.~M.~M., C.~T.~K and S.~G., with inputs from all co-authors. S.~G. supervised the experimental work in Delft.\\

\noindent
\textbf{Notes}\\
The authors declare no competing interests.\\

\noindent
\textbf{\large Acknowledgments}\\
We thank Michael Wimmer and Michal P. Nowak for useful discussions. Moreover we would like to thank J.H. Dycus, M.E. Salmon and R.E. Daniel from Eurofins EAG Materials Science for the TEM studies. The research at Delft was supported by the Dutch National Science Foundation (NWO), the Early Research Programme of the Netherlands Organisation for Applied Scientific Research (TNO) and a TKI grant of the Dutch Topsectoren Program. The work at Purdue was funded by Microsoft Quantum.\\

\noindent
\textbf{\large References}\\

\clearpage

\newcommand{\beginsupplement}{%
	\setcounter{section}{0}
	\renewcommand{\thesection}{S\arabic{section}}%
	\setcounter{table}{0}
	\renewcommand{\thetable}{S\arabic{table}}%
	\setcounter{figure}{0}
	\renewcommand{\thefigure}{S\arabic{figure}}%
	\setcounter{equation}{0}
	\renewcommand{\theequation}{S\arabic{equation}}%
}
\onecolumngrid

\section*{\large{Supplementary Information}}
\vspace{10mm}

\beginsupplement

\section{Wafer growth and characterization}

\begin{figure}[!ht]
	\includegraphics[width = 0.8\textwidth]{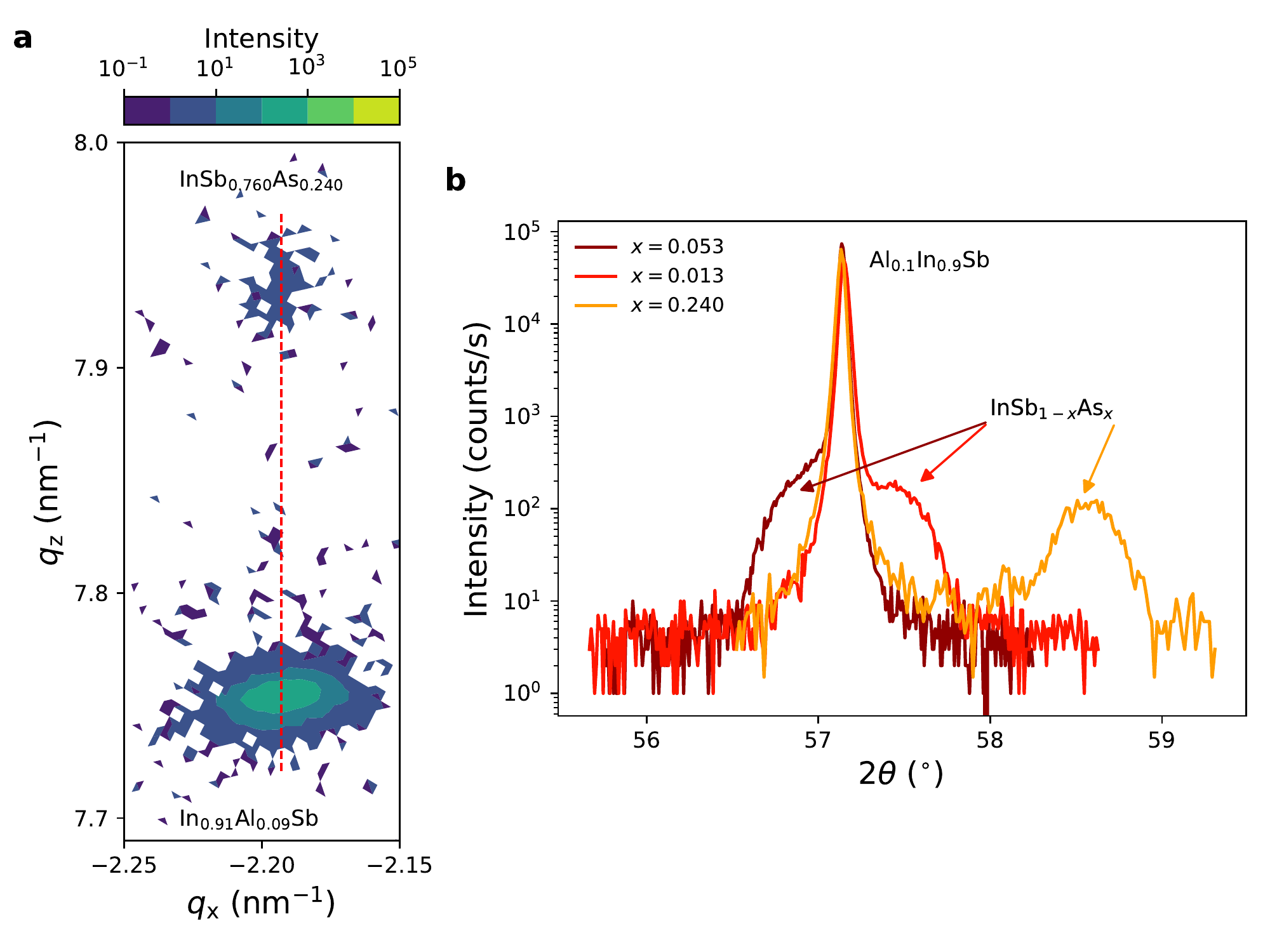}
	\caption{\textbf{a,} Asymmetric (115) reciprocal space map for the $\mathrm{InSb_{0.760}As_{0.240}}$ quantum well heterostructure. \textbf{b,} High-resolution x-ray diffraction $2\theta-\omega$ scans obtained along (004) for the $\mathrm{InSb}_{1-x}\mathrm{As}_{x}$ quantum well heterostructures with different As concentrations.}
	\label{supp_fig0}
\end{figure}

InSbAs quantum well heterostructures are grown in a Veeco Gen 930 molecular beam epitaxy (MBE) system equipped with valved crackers for As and Sb, and effusion cells for Ga, In, and Al. The growths are performed on GaAs (100) substrates with a \SI{0.5}{\degree} miscut toward the (111)B crystalline direction to minimize surface roughness~\cite{Shi_2019_S}. The growth is initiated with a \SI{100}{nm} thick GaAs buffer layer, directly followed by a \SI{1}{\mu m} thick AlSb layer and a \SI{4}{\mu m} thick $\mathrm{Al_{0.1}In_{0.9}Sb}$ layer that also serves as bottom barrier for the InSbAs quantum wells. The AlSb and AlInSb layers help accommodate the large lattice mismatch between the GaAs substrate and InSbAs by forming a pseudo-substrate on which InSbAs can be coherently strained. 

The growth rate for all the semiconducting layers is kept at roughly \SI{0.5}{ML/s}, under group V element rich conditions. The incorporation of As and the resulting alloy content of the InSbAs quantum well is controlled by the growth temperature and the Sb/As flux ratio. The growth temperature is monitored by a BandiT spectrometer through blackbody radiation fitting. For the $x=0.053$, 0.130 and 0.240 samples, the growth temperatures were \SI{445}{\degreeCelsius}, \SI{477}{\degreeCelsius} and \SI{483}{\degreeCelsius}, respectively. While the Sb flux was kept around \SI{4.2e-7}{Torr} for all three samples to maintain group V overpressure, the As flux changed from \SI{4.2e-6}{Torr} ($x=0.053$ and 0.130) to \SI{1.1e-5}{Torr} ($x=0.240$). The semiconductor growth terminates under As rich conditions with the epitaxy of 2 ML InAs to prevent interdiffusion between the quantum well and the superconducting layer on top~\cite{Thomas_2019_S}. 

The heterostructures are then transferred under ultra-high vacuum to a Veeco 620 MBE system equipped with an Al effusion cell, and a quartz crystal monitor to determine the growth rate. Inside this MBE chamber, a moveable cryocooler is used to contact and cool the wafers down to liquid nitrogen temperature within a few hours~\cite{Thomas_2019_S}. After 6 hours of cooling, a 7 nm thin layer of Al is deposited on the semiconductor heterostructure with a typical growth rate of \SI{0.3}{\AA /s}. Immediately after the Al deposition, the samples are transferred into another chamber where they are oxidized for 15 min under an $\mathrm{O_2}$ pressure of \SI{5e-5}{Torr} to stabilize the Al films~\cite{Gazibegovic_2017_S}.

High-resolution X-ray diffraction (HRXRD) measurements were performed to evaluate the As concentration, $x$, of the $\mathrm{InSb}_{1-x}\mathrm{As}_{x}$ quantum wells and reciprocal space maps were acquired to assess the strain of these layers. These measurements were performed using a X’pert PANalytical diffractometer with a copper X-ray tube operating at a wavelength $\lambda=\SI{1.5406}{\AA}$. Asymmetric (115) reciprocal space maps (averaged at 2 opposite azimuth angles $\phi=\SI{0}{\degree}$ and \SI{180}{\degree}) confirm that the $\mathrm{InSb}_{1-x}\mathrm{As}_{x}$ layers are coherently strained for the largest investigated composition $x=0.240$, as shown in Fig. \ref{supp_fig0}a, where the InSbAs peak is along the same red dashed vertical line as the \SI{100}{\percent} relaxed AlInSb peak. Similar reciprocal space maps were obtained for the heterostructures with $x=0.053$ and 0.130, confirming that InSbAs quantum wells are coherently and fully strained in the investigated composition range. 
$2\theta-\omega$ scans are presented in \ref{supp_fig0}b. The peak positions of the coherently strained $\mathrm{InSb}_{1-x}\mathrm{As}_{x}$ layers give the As concentration for each sample, using the 4 micron fully relaxed $\mathrm{Al_{0.1}In_{0.9}Sb}$ layer as the substrate in the analysis.

\section{Device fabrication}
The processing of Sb-based 2DEGs in proximity to Aluminum is more challenging than InAs/Al systems. This is due to the potential intermixing of Al and Sb, which becomes more severe at elevated temperatures. We therefore need to ensure that all the processing steps (as described below) are performed at as low a temperature as possible. This includes room temperature “baking” of the resist for every lithography step, and low temperature atomic layer deposition.    

The devices presented in the main text are fabricated using electron beam lithography. First, the Al and the 2DEG is etched in unwanted areas. The Al etch is performed in Transene D etchant at a temperature of \SI{48.2}{\celsius} for \SI{9}{s}. Subsequently, using the same PMMA mask, the 2DEG is etched in a wet etch solution consisting of \SI{560}{ml} deionized water, \SI{9.6}{g} citric acid powder, \SI{5}{ml} $\text{H}_{2}\text{O}_{2}$ and \SI{4}{ml} $\text{H}_3\text{PO}_4$. We etch for \SI{2}{min}, resulting in an etch depth around \SI{100}{nm}. A second Al etch is performed to define the normal regions using Transene D etchant at a temperature of \SI{38.2}{\celsius} for \SI{17}{s}. Next, a \SI{60}{nm} thick $\mathrm{SiN}_x$ dielectric layer is sputtered for the Hall bars and the Josephson junction/spectroscopy devices. For the superconducting island and the quasi-1D superconducting strip we use a \SI{40}{nm} thick $\mathrm{AlO}_x$ dielectric deposited by atomic layer deposition at \SI{40}{\degreeCelsius}. Finally, top-gates are deposited by evaporating \SI{10}{nm}/\SI{190}{nm} of Ti/Au. 

\section{Mobility and density}

\begin{figure}[!h]
	\includegraphics[width=1.0\textwidth]{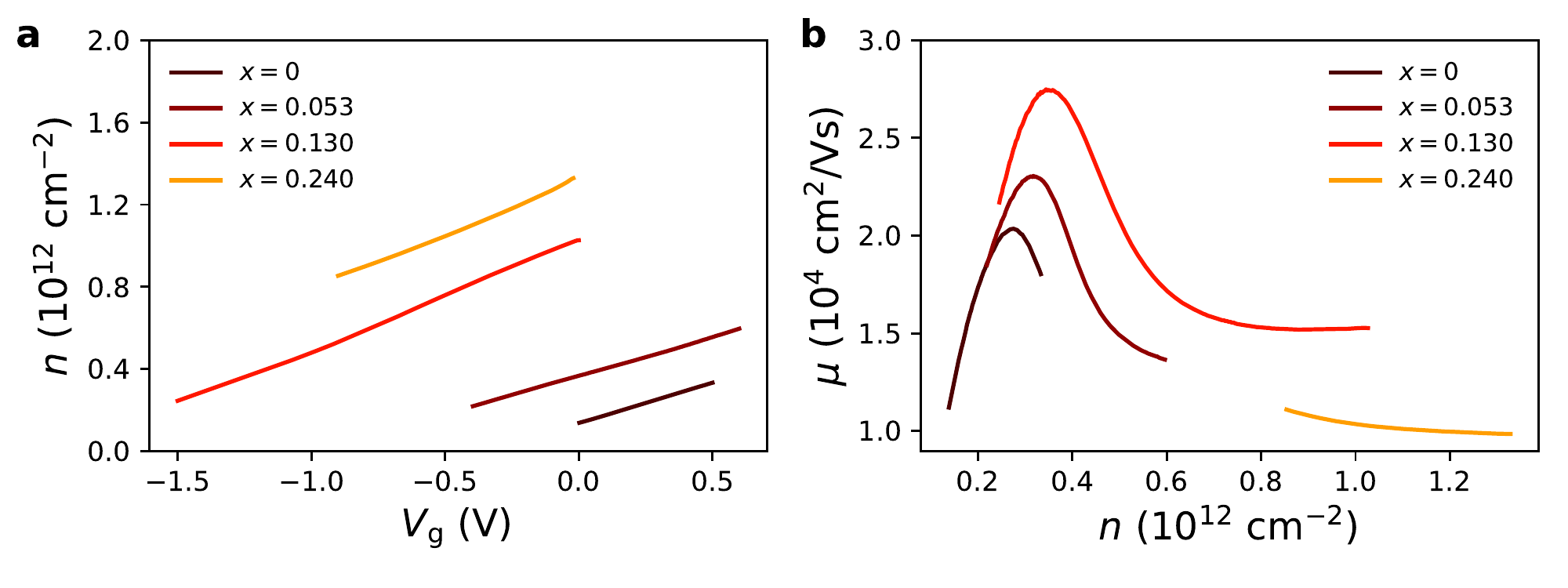}
	\caption{\textbf{a,} Electron density plotted against the gate voltage for all As concentrations. At a fixed gate voltage, the electron density increases monotonically with $x$. \textbf{b,} Electron mobility as a function of $n$.}
	\label{supp_fig1}
\end{figure} 

The $\mathrm{InSb}_{1-x}\mathrm{As}_{x}$ 2DEGs are characterized by measuring the Hall effect in a gated Hall bar geometry at $T=\SI{300}{mK}$. From a linear fit to the transversal resistance in a magnetic field range up to $\pm\SI{0.5}{T}$, we extract the electron density, $n$, at every gate voltage, $V_{\mathrm{g}}$. Figure \ref{supp_fig1}a shows $n$ plotted against $V_{\mathrm{g}}$ for all As concentrations, $x$. At a fixed gate voltage, the electron density increases systematically with $x$. The nominal density, $n (V_{\mathrm{g}}=\SI{0}{V})$, increases from $\SI{0.15e12}{cm^{-2}}$ for $x=0$, to $\SI{1.35e12}{cm^{-2}}$ for $x=0.240$. This suggests that the incorporation of As causes an increasing negative band offset at the dielectric-2DEG interface. 

Using the longitudinal resistivity, we calculate the mobility, $\mu$, at every gate voltage. Figure \ref{supp_fig1}b shows a plot of $\mu$ as a function of $n$ for all As concentrations. We observe peak mobilities between $\SI{20000}{cm^{2}/Vs}$ and $\SI{28000}{cm^{2}/Vs}$ (the peak mobility for $x=0.240$ could not be reached because of gate leakage). The peak mobilities occur around $n=\SI{0.35e12}{cm^{-2}}$, with a slight trend of moving towards higher densities for higher As concentrations. We attribute the peak mobilities to the de-population of the second subband. 

\section{Effect of electron density on spin-orbit coupling}

\begin{figure}[!h]
	\includegraphics[width=1.0\textwidth]{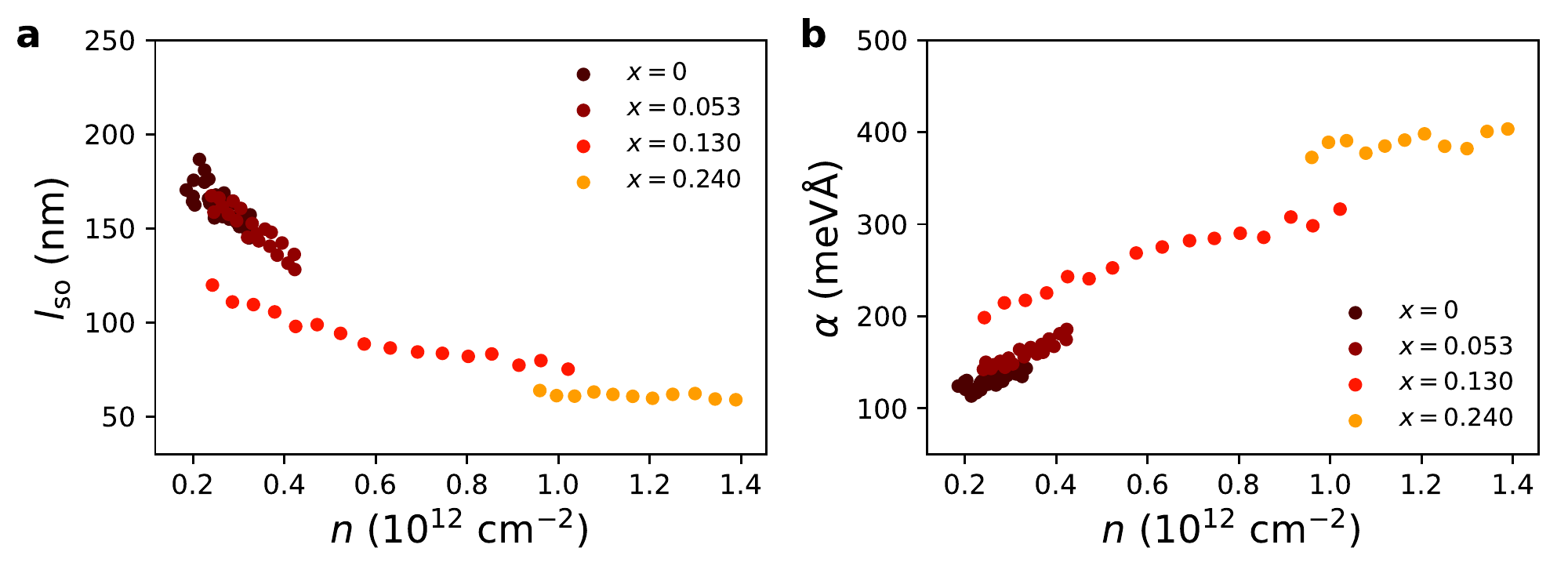}
	\caption{\textbf{a,} Spin-orbit length plotted against the electron density for all As concentrations. \textbf{b,} Linear Rashba parameter plotted against $n$. When comparing at the same density, $l_{\mathrm{so}}$ ($\alpha$) decreases (increases) for increasing As concentration.}
	\label{supp_fig2}
\end{figure}

In Fig. 2 of the main text we show the spin-orbit length, $l_{\mathrm{so}}$, and the linear Rashba paramter, $\alpha$, plotted against the gate voltage for all As concentrations, $x$. In Fig. \ref{supp_fig2}a and b we show $l_{\mathrm{so}}$ and $\alpha$ plotted against the electron density, $n$, respectively. When comparing at the same density, the overall trend of decreasing (increasing) $l_{\mathrm{so}}$ ($\alpha$) with increasing As concentration persists.

\section{Effective mass and g-factor}

\begin{figure*}[!h]
	\includegraphics[width=1.0\textwidth]{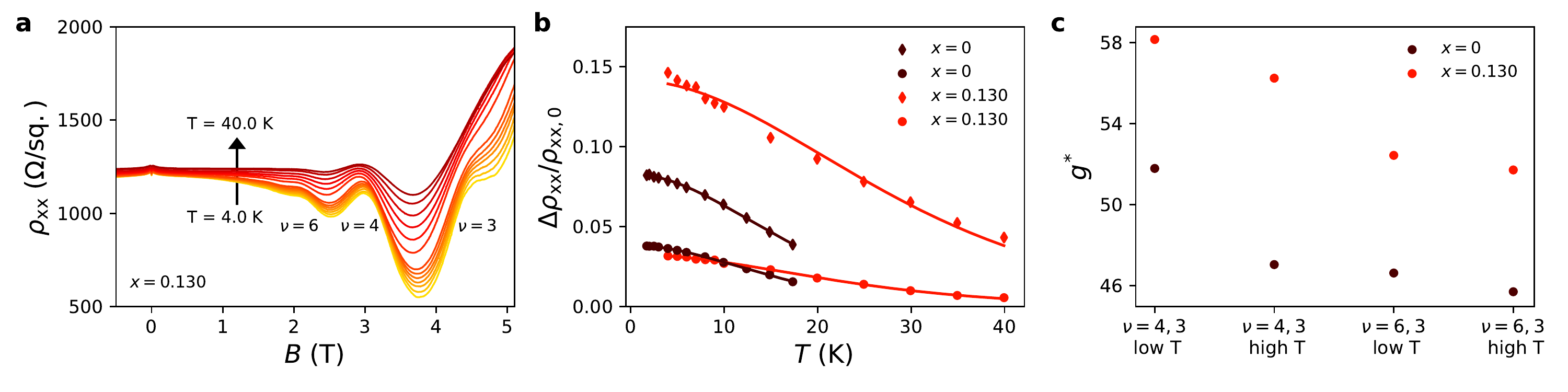}\\
	\caption{\textbf{a,} Shubnikov-de Haas oscillations for temperatures ranging from \SI{4}{K} to \SI{40}{K} for the $x=0.130$ sample at a fixed density of $n=\SI{3.41e11}{cm^{-2}}$ close to peak mobility. \textbf{b,} Temperature dependence of the oscillation amplitude for the $\nu=6$ minima (circles) and maxima (rhombs) for $x=0$ and $x=0.130$. The solid lines are fits to the data in order to obtain the effective mass. \textbf{c,} g-factor, $g^*$, for $x=0$ and $x=0.130$ considering the even-odd filling factor couple $\nu=4-3$ and $\nu=6-3$. The fits are performed in a high and low temperature range.}
	\label{supp_fig3}
\end{figure*}

We extract the effective mass, $m^*$, of the electrons in the $\mathrm{InSb}_{1-x}\mathrm{As}_{x}$ 2DEGs by measuring the temperature dependence of the Shubnikov-de Haas (SdH) oscillations amplitude in gated Hall bars. Figure \ref{supp_fig3}a shows the longitudinal resistivity, $\rho_{\mathrm{xx}}$, as a function of perpendicular magnetic field, $B$, for $x=0.130$ at a fixed density of $n=\SI{3.41e11}{cm^{-2}}$, close to peak mobility. The temperature ranges from $T=$ \SIrange{4}{40}{K}. The same measurement is also done for $x=0$ at $n=\SI{2.93e11}{cm^{-2}}$ (for $T=$ \SIrange{1.7}{17.3}{K}). After a polynomial background subtraction, the effective mass can be obtained from a fit to the thermal damping of the SdH oscillation amplitude, $\Delta \rho_{\mathrm{xx}}$, normalized to the zero-field magnetoresistance value, $\rho_{xx,0}(T)$, at a fixed filling factor, $\nu$, using~\cite{Lange_1993_S}:
\begin{equation*}
	\frac{\Delta \rho_{\mathrm{xx}}(T)}{\rho_{\mathrm{xx, 0}}(T)}\propto\frac{\alpha T}{\sinh(\alpha T)},
\end{equation*}
Here, $\alpha=\pi k_{\mathrm{B}}m^{*}\nu/(\hbar^{2}n)$, where $k_{\mathrm{B}}$ is the Boltzmann constant and $\hbar$ is the Planck constant. Figure \ref{supp_fig3}b shows such fits for the $\nu=6$ minima (circles) and maxima (rhombs) for $x=0$ and $x=0.130$. The resulting effective mass is lower for the ternary sample, with a weighted mean value of $m^*=(0.0162\pm 0.0004)m_0$, where $m_0$ is the free electron mass. The $x=0$ sample shows a heavier effective mass of $m^*=(0.0180\pm 0.0002)m_0$.

We proceed by extracting the g-factor, $g^*$, from the temperature dependence of the SdH oscillations, extending the temperature range from $T=$ \SIrange{1.7}{44.7}{K}. Using the method reported in~\cite{Drichko_2018_S}, an expression for $g^*$ can be obtained by combining the equations for the thermal activation energy of an even-odd filling factor couple. We consider the filling factor couples $\nu=4-3$ and $\nu=6-3$. Since we find different activation energies from the linear fit of $\ln(\rho_{xx})$ vs. $1/T$ depending on the temperature range, we report in Fig. \ref{supp_fig3}c the extracted values for the low- and high-temperature ranges. The $m^*$ values we found earlier are used in the $g^*$ calculation. We find an average g-factor of $g^*=47.8\pm 2.8$ for $x=0$, and $g^*=54.6\pm 3.1$ for $x=0.130$. 

\section{Multiple Andreev reflections}

\begin{figure}[!h]
	\includegraphics[width = 1.0\textwidth]{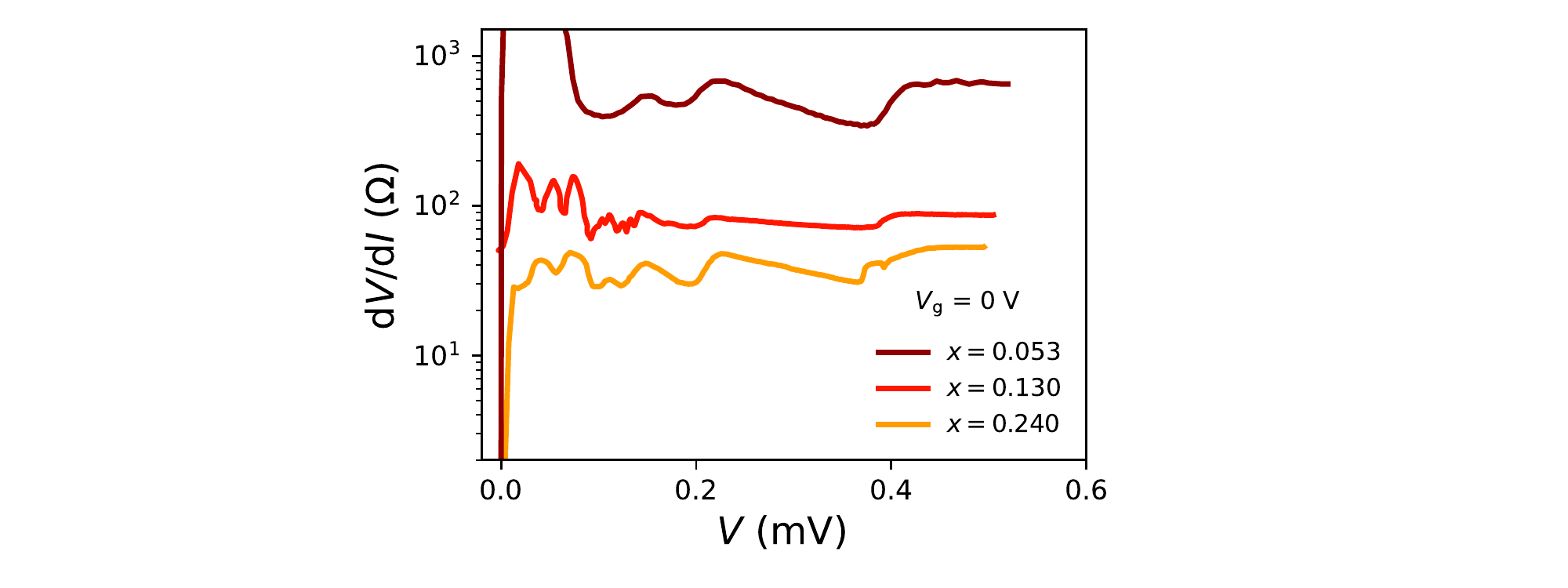}
	\caption{Differential resistance plotted against the voltage drop between the two Al leads for the JJs with $x=0.053$, 0.130 and 0.240. All measurements are obtained at a JJ gate voltage of $V_{\mathrm{g}}=\SI{0}{V}$. The curve for $x=0.130$ is offsetted for clarity. MAR up to several orders are observed for all As concentrations.}
	\label{sup_fig_MAR}
\end{figure}

In Fig. 3b and c of the main text we show a representative Fraunhofer interference pattern and a gate dependence of the switching current for a Josephson junction with $x=0.053$. As demonstrated in Fig. \ref{sup_fig_MAR}, we observe pronounced subgap conductance modulations that are due to multiple Andreev reflections (MAR) for all As concentrations. The observation of higher order MAR shows that transport is phase-coherent across a length scale of several times the junction length.


%

\end{document}